\documentclass[twoside,twocolumn,9pt]{article}
\usepackage{extsizes}
\usepackage[super,sort&compress,comma]{natbib} 
\usepackage[version=3]{mhchem}
\usepackage[left=1.5cm, right=1.5cm, top=1.785cm, bottom=2.0cm]{geometry}
\usepackage{balance}
\usepackage{mathptmx}
\usepackage{sectsty}
\usepackage{graphicx} 
\usepackage{lastpage}
\usepackage[format=plain,justification=justified,singlelinecheck=false,font={stretch=1.125,small,sf},labelfont=bf,labelsep=space]{caption}
\usepackage{float}
\usepackage{fancyhdr}
\usepackage{fnpos}
\usepackage[english]{babel}
\addto{\captionsenglish}{%
  
}
\usepackage{array}
\usepackage{droidsans}
\usepackage{charter}
\usepackage[T1]{fontenc}
\usepackage[usenames,dvipsnames]{xcolor}
\usepackage{setspace}
\usepackage[compact]{titlesec}
\usepackage{hyperref}

\usepackage{epstopdf}

\usepackage{tikz}
\usetikzlibrary{shapes}

\definecolor{cream}{RGB}{222,217,201}

\begin{document}

\pagestyle{fancy}
\thispagestyle{plain}
\fancypagestyle{plain}{
\renewcommand{\headrulewidth}{0pt}
}

\makeFNbottom
\makeatletter
\renewcommand\LARGE{\@setfontsize\LARGE{15pt}{17}}
\renewcommand\Large{\@setfontsize\Large{12pt}{14}}
\renewcommand\large{\@setfontsize\large{10pt}{12}}
\renewcommand\footnotesize{\@setfontsize\footnotesize{7pt}{10}}
\makeatother

\renewcommand{\thefootnote}{\fnsymbol{footnote}}
\renewcommand\footnoterule{\vspace*{1pt}%
\color{cream}\hrule width 3.5in height 0.4pt \color{black}\vspace*{5pt}} 
\setcounter{secnumdepth}{5}

\makeatletter 
\renewcommand\@biblabel[1]{#1}            
\renewcommand\@makefntext[1]%
{\noindent\makebox[0pt][r]{\@thefnmark\,}#1}
\makeatother 
\renewcommand{\figurename}{\small{Fig.}~}
\sectionfont{\sffamily\Large}
\subsectionfont{\normalsize}
\subsubsectionfont{\bf}
\setstretch{1.125} 
\setlength{\skip\footins}{0.8cm}
\setlength{\footnotesep}{0.25cm}
\setlength{\jot}{10pt}
\titlespacing*{\section}{0pt}{4pt}{4pt}
\titlespacing*{\subsection}{0pt}{15pt}{1pt}

\fancyfoot{}
\fancyfoot[LO,RE]{\vspace{-7.1pt}\includegraphics[height=9pt]{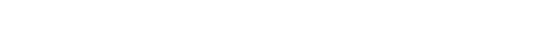}}
\fancyfoot[CO]{\vspace{-7.1pt}\hspace{13.2cm}\includegraphics{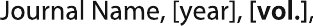}}
\fancyfoot[CE]{\vspace{-7.2pt}\hspace{-14.2cm}\includegraphics{RF}}
\fancyfoot[RO]{\footnotesize{\sffamily{1--\pageref{LastPage} ~\textbar  \hspace{2pt}\thepage}}}
\fancyfoot[LE]{\footnotesize{\sffamily{\thepage~\textbar\hspace{3.45cm} 1--\pageref{LastPage}}}}
\fancyhead{}
\renewcommand{\headrulewidth}{0pt} 
\renewcommand{\footrulewidth}{0pt}
\setlength{\arrayrulewidth}{1pt}
\setlength{\columnsep}{6.5mm}
\setlength\bibsep{1pt}

\makeatletter 
\newlength{\figrulesep} 
\setlength{\figrulesep}{0.5\textfloatsep} 

\newcommand{\topfigrule}{\vspace*{-1pt}%
\noindent{\color{cream}\rule[-\figrulesep]{\columnwidth}{1.5pt}} }

\newcommand{\botfigrule}{\vspace*{-2pt}%
\noindent{\color{cream}\rule[\figrulesep]{\columnwidth}{1.5pt}} }

\newcommand{\dblfigrule}{\vspace*{-1pt}%
\noindent{\color{cream}\rule[-\figrulesep]{\textwidth}{1.5pt}} }

\makeatother

\twocolumn[
  \begin{@twocolumnfalse}
{\includegraphics[height=30pt]{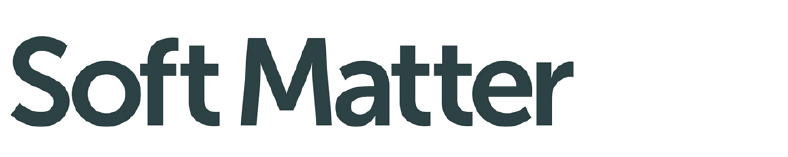}\hfill\raisebox{0pt}[0pt][0pt]{\includegraphics[height=55pt]{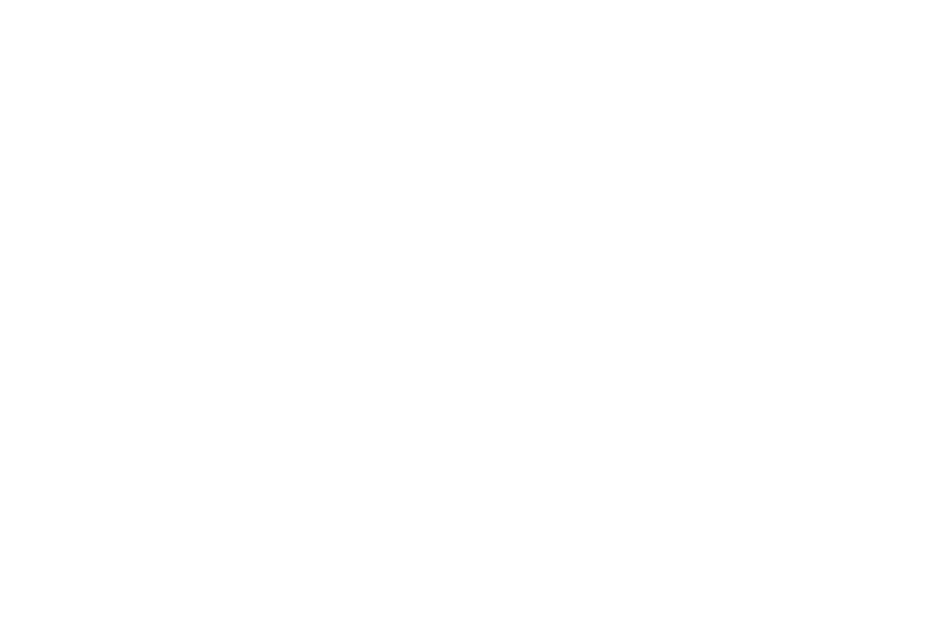}}\\[1ex]
\includegraphics[width=18.5cm]{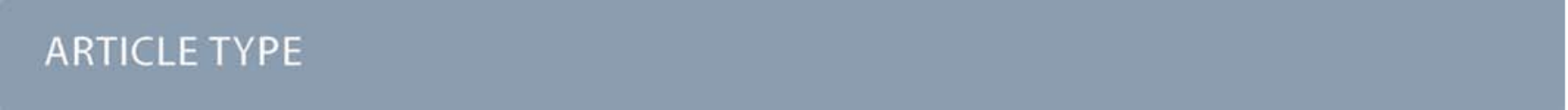}}\par
\vspace{1em}
\sffamily
\begin{tabular}{m{4.5cm} p{13.5cm} }

\includegraphics{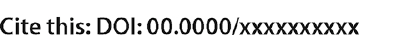} & \noindent\LARGE{\textbf{Mechanical properties of model colloidal mono-crystals$^\dag$}} \\
\vspace{0.3cm} & \vspace{0.3cm} \\

 & \noindent\large{Jean-Christophe~Ono-dit-Biot,\textit{$^{a}$} Pierre~Soulard,\textit{$^{b}$} Solomon~Barkley,\textit{$^{a}$} Eric~R.~Weeks,\textit{$^{c}$} Thomas~Salez,\textit{$^{d,e}$} Elie~Rapha\"el,\textit{$^{b}$}  and Kari~Dalnoki-Veress\textit{$^{\ast}$$^{a,b}$}} \\


\includegraphics{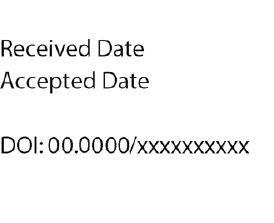} & \noindent\normalsize{We investigate the elastic and yielding properties of  two dimensional defect-free mono-crystals made of highly monodisperse droplets. Crystals are compressed between two parallel boundaries of which one acts as a force sensor. As the available space between boundaries is reduced, the crystal goes through successive row-reduction transitions.  For small compression forces, the crystal responds elastically until a critical force is reached and the assembly fractures in a single catastrophic global event. Correspondingly there is a peak in the force measurement associated with each row-reduction. The elastic properties of ideal mono-crystal samples are fully captured by a simple analytical model consisting of an assembly of individual capillary springs. The yielding properties of the crystal are captured with  a minimal bond breaking model.} \\

\end{tabular}

 \end{@twocolumnfalse} \vspace{0.6cm}

  ]

\renewcommand*\rmdefault{bch}\normalfont\upshape
\rmfamily
\section*{}
\vspace{-1cm}
\newcommand{\markercirc}{\raisebox{0.5pt}{\tikz{\node[draw,scale=0.4,circle,fill=black](){};}}}
\newcommand{\markersquare}{\raisebox{0.5pt}{\tikz{\node[draw,scale=0.4,regular polygon, regular polygon sides=4,fill=black](){};}}}
\newcommand{\markerstar}{\raisebox{0.5pt}{\tikz{\node[draw,scale=0.4,star, star points =5, star point ratio=0.3, fill=black,rotate=180](){};}}}
\newcommand{\markerdiamond}{\tikz{\node[draw,scale=0.4,star,diamond, fill=black](){};}}
\newcommand{\markertriangle}{\tikz{\node[draw,scale=0.4,regular polygon, regular polygon sides=3,rotate=180,fill=black](){};}}

\footnotetext{\textit{$^{\ast}$}~dalnoki@mcmaster.ca}
\footnotetext{\textit{$^{a}$}~Department of Physics \& Astronomy, McMaster University, Hamilton, ON, L8S 4L8, Canada.}
\footnotetext{\textit{$^{b}$}~UMR CNRS Gulliver 7083, ESPCI Paris, PSL Research University, 10 rue Vauquelin, 75005 Paris, France. }
\footnotetext{\textit{$^{c}$}~Department of Physics, Emory University, Atlanta, GA 30322, USA. }
\footnotetext{\textit{$^{d}$}~Univ. Bordeaux, CNRS, LOMA, UMR 5798, F-33405 Talence, France. }
\footnotetext{\textit{$^{e}$}~Global Station for Soft Matter, Global Institution for Collaborative Research and Education, Hokkaido University, Sapporo, Japan}

\footnotetext{\dag~Electronic Supplementary Information (ESI) available. See DOI: 10.1039/cXsm00000x/}



\section{Introduction}

Historically, foams have often been used as model materials, with an especially inspiring example being the use of bubble rafts to model the behaviour of a metallic structure proposed by Bragg and Nye~\cite{bragg1947}.  Using bubbles instead of atoms, dislocations and grain boundaries were imaged directly and mechanical properties of the assembly were studied~\cite{bragg1949}. The use of foams, emulsions, and colloids has become a powerful tool to study fundamental questions such as the glass transition~\cite{pusey1987, weeks2000, hunter2012, manoharan2015,Illing2016,Vivek2017}, formation and melting of crystals~\cite{schaefer1975, Pusey1986,alsayed2005,wang2012}, the order-to-disorder transition~\cite{Yunker2010,Hanifpour2014,Goodrich2014,Keim2015,charbonneau2019,ono2020} and jamming~\cite{liu1998, trappe2001, Ohern2003, corwin2005,liu2010,bi2011}. Complex biological systems can also be modeled using foams and emulsions~\cite{hayashi2004, gonzalez2012, pontani2012, douezan2012}. One of the important characteristics of these model systems is their mechanical response to external stress such as compression or shear. The mechanical properties of such systems depend strongly on the volume fraction of suspended particles~\cite{princen1983,mason1995} as well as the interaction between particles~\cite{irani2014,grob2014,Bonn2017,golovkova2020}. Above a critical volume fraction, foams and emulsions behave as soft solids~\cite{mason1996, Coussot2002,Goyon2008,Dollet2014,Bonn2017}.
For small applied stress, the assembly deforms elastically~\cite{bragg1949,mason1995}. When the magnitude of the applied stress exceeds a critical value, given by the yield stress, plastic deformations occur and the material flows as a liquid. Several theoretical~\cite{picard2004, bocquet2009, kamrin2012, nicolas2013} and simulation~\cite{kabla2003,tewari2009,mansard2013,Dollet2015} works have studied the connection between local plastic events and macroscopic flow. As the particles constituting the foam or the emulsion can be resolved individually~\cite{weeks2000}, studies linking microscopic plastic events to flow properties can also be conducted experimentally~\cite{Schall2007,jop2012, knowlton2014, Dollet2015, Chen2015, Gai2016,golovkova2020}. 



In foams and emulsions, the nature of the constituting particles is also a key parameter in understanding the properties of the assembly. For example, the relevant scale for the elastic modulus of an assembly of oil droplets is set by the Laplace pressure~\cite{mason1995}. Thus changing the size of the droplets or the interfacial tension modifies the elastic properties of the structure. The size distribution of the particles is also particularly important. Indeed, monodisperse particles can assemble into a crystal while polydispersity prevents crystallization~\cite{Pusey1987poly,Auer2001}. Due to their perfect arrangement and periodicity, defect-free monodisperse crystals are well understood theoretically~\cite{princen1983}. However, these mono-crystals are more challenging to study experimentally, as perfect monodispersity and crystalline order are difficult to achieve. Most experimental studies on colloidal crystals have focused on the study of polycrystals~\cite{mason1996,van2006,Gai2016,golovkova2020} and in particular plastic deformations resulting from shear imposed on structurally disordered materials~\cite{schall2004,vecchiolla2019}. For crystals, it is known that the mechanical properties, and in particular the yield stress, is dictated by dislocations or local structural disorder~\cite{kittel1976}. To date, only a small number of experimental studies have been able to produce ideal defect-free mono-crystals~\cite{rosa1998, gouldstone2001} and the study of their elastic and yielding properties warrants further attention.

In this study we use lightly attractive oil droplets in water with low  polydispersity to create mono-crystals made of tens of droplets. Due to the droplets being monodisperse combined with a small sample size, the aggregates are defect-free crystals. In the experiments we simultaneously measure the mechanical response of these ideal mono-crystals under compression between two glass capillaries and image the rearrangements that cause yielding and plastic deformation of the structure. We find that crystals behave elastically until a critical force is applied and the crystal fractures. The bonds between droplets are broken in a coordinated manner, after which the aggregate can no longer sustain any stress. Upon further compression the structure rearranges into a new crystal with one less row of droplets. The elastic response and the yield properties are fully captured by a simple analytical model. 

\section{Experimental Methods}

\begin{figure}
\centering
  \includegraphics[width=.9\columnwidth]{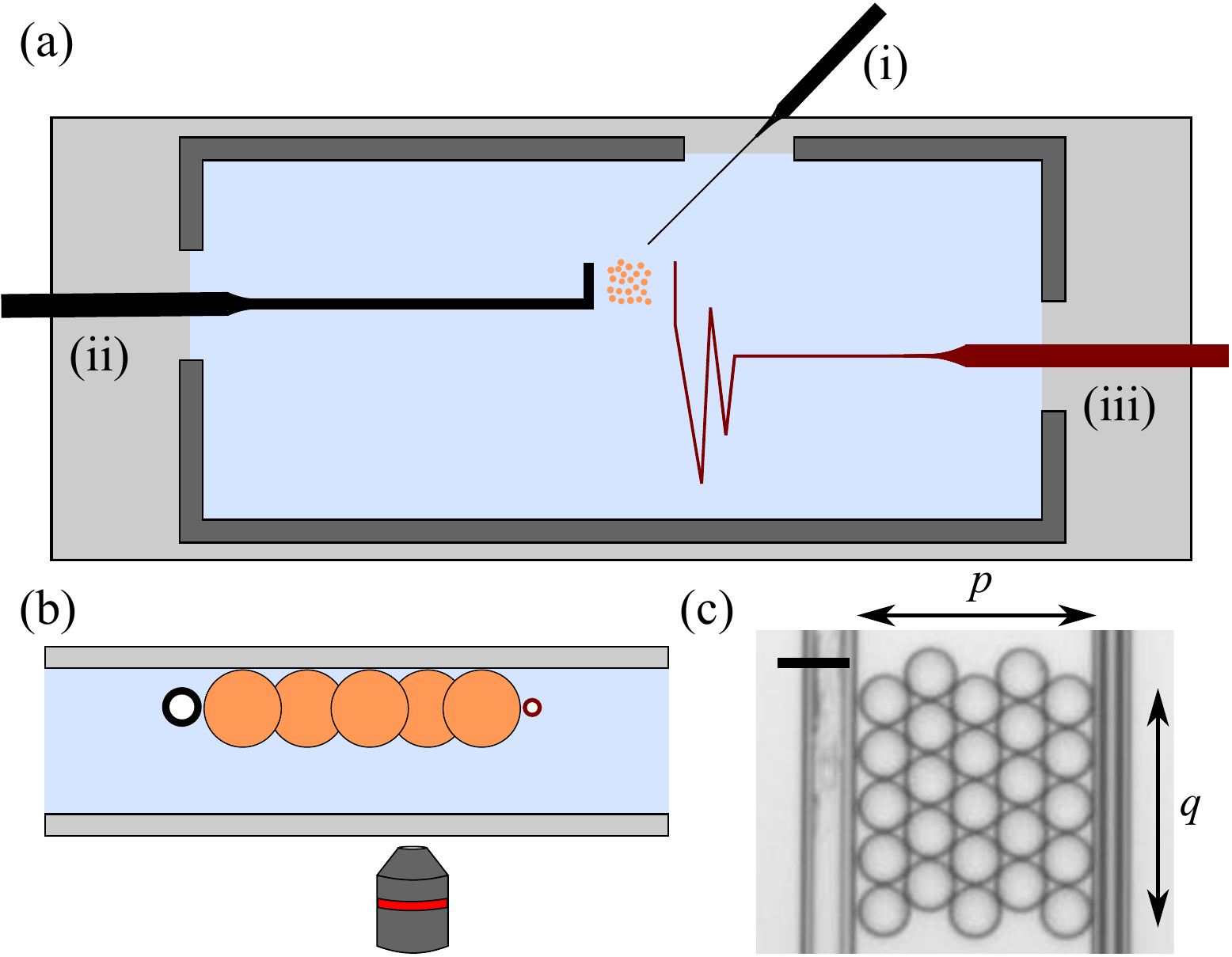}
  \caption{\label{fig1}(a) Schematic top view of the experimental chamber. The typical dimensions of the wall (dark grey) are $55 \times 30 \times 2.5$~mm$^3$.  The ``droplet pipette", ``pushing pipette", and ``force-sensing pipette" are labelled as (i), (ii) and (iii) respectively.  (b) Schematic side view (not to scale). The buoyant droplets are assembled into a quasi 2D crystal under the top glass plate. The pushing pipette (black circle) is moved at speed $v=0.3$ $\mu$m.s$^{-1}$ to compress the aggregate and the force-sensing pipette (red circle) is used to measure forces. Both pipettes are placed near the equatorial plane of the droplets so forces are applied horizontally. (c) Optical microscopy image of a typical crystal (scale bar is 50~$\mu$m). $p$ refers to the number of rows of droplets and $q$ the number of droplets per row. In this example $p=q=5$.}

\end{figure}

The experimental chamber ($55 \times 30$~mm$^2$), shown in Fig.~\ref{fig1}, is made of two glass slides separated by a 3D-printed spacer of $2.5$~mm (not shown). This gap between the glass slides is $10^3$ times larger than the size of the droplets. The chamber is filled with an aqueous solution with 3\% (w/w) sodium dodecyl sulfate (SDS) and 1.5\% (w/w) NaCl. The 3D-printed wall reduces evaporation of the solution and ensures a concentration that is approximately constant over the course of the experiments. At this concentration, the surfactant, SDS, assembles into micelles acting as a depletant resulting in a short-range attraction between the droplets~\cite{Bibette1992}. Glass capillaries (World Precision Instruments, USA) are pulled to a diameter of about 10 $\mu$m over several centimeters in length using a pipette puller (Narishige, Japan). One of these pipettes, the ``droplet pipette'' is used to produce highly monodisperse droplets of mineral oil, with size proportional to the tip radius of the pipette, using the snap-off instability~\cite{Barkley2016}. The droplets used in this experiment have a radius $R=18.9 \pm 0.3 \ \mu$m. The uncertainty on the radius corresponds to the precision on the measurement of the droplet size. As droplets are produced using the snap-off instability, the droplet polydispersity is less than 0.7\%~\cite{Barkley2016}.
Droplets are buoyant and accumulate under the top glass slide. Aggregates of oil droplets are assembled droplet-by-droplet into 2D crystals with arbitrary shapes (see Movie M1 in Supplemental Material). The crystals are made up of $p$ rows and $q$ droplets per row, with the initial aggregate defined as $p=p_\mathrm{ini}$ and $q=q_\mathrm{ini}$ [see  Fig.~\ref{fig1}~(c)]. Under compression, the crystal rearranges with corresponding values of $p$ and $q$, while keeping the total number of droplets, $N_\mathrm{tot}$, constant. Aggregates are compressed between two micropipettes: the ``pushing pipette'' and the ``force-sensing pipette''. The ``pushing pipette'' is a short and stiff pipette used to compress the aggregates. The pushing pipette is affixed to a motorized translation stage and moved at a constant speed, $v=0.3$~$\mu$m.s$^{-1}$, for all experiments. The ``force-sensing pipette'' is a long compliant pipette. Its deflection is used to measure the forces applied to the aggregate as it is compressed~\cite{Backholm2019}. The pipette is pulled to a diameter of $\sim 10$~$\mu$m over a length of $\sim 3$~cm to be sensitive to forces as small as $\sim 100$~pN. The thin section of the pipette is locally and temporarily heated to soften the glass such that the pipette can be bent into a shape that fits in the small chamber while maximizing its total length [see pipette (iii) in Fig.~\ref{fig1}~(a)]. The pushing and force-sensing pipettes are aligned to be as parallel as possible in order to compress the aggregate uniformly.  A misalignment would result in one side of the aggregate breaking earlier than the other. The chamber is placed atop of an inverted optical microscope for imaging while the aggregates are compressed and images are collected at a frame rate of 1.8 Hz.

The distance between the pushing pipette and the force-sensing pipette, $\delta$, is measured using cross-correlation analysis between images. This analysis leads to a sub-pixel resolution and in this study a precision of $\sim 0.1$~$\mu$m \cite{Backholm2019}. The deflection of the pipette is measured using the cross-correlation analysis and converted into a  force using the calibrated spring constant $k_\mathrm{p}=1.3\pm 0.1$~nN.$\mu$m$^{-1}$ of the force-sensing pipette~\cite{Backholm2019}. 
The crystal is fractured and rearranges under compression by breaking adhesive bonds between droplets. Using the optical microscopy images, fracture events observed directly can be linked to features in the measured force-distance curves.

\section{Results and Discussion}

\subsection{Compression of colloidal crystals}

\begin{figure}[h!]
\centering
  \includegraphics[width=.9\columnwidth]{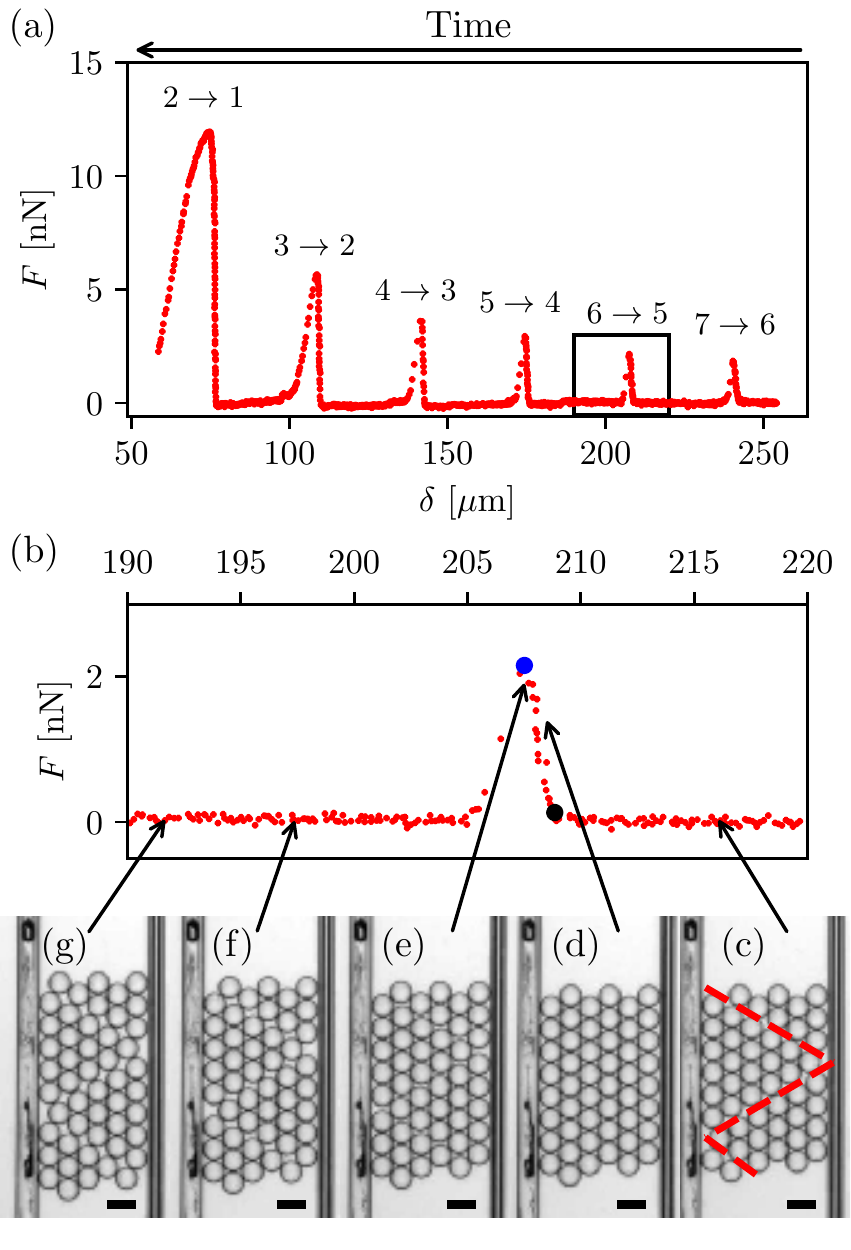}
  \caption{(a) Measured force, $F$, as a function of the distance between the pipettes, $\delta$ for a crystal with an initial configuration given by $p_\mathrm{ini}$=7; $q_\mathrm{ini}$=7, and a droplet size $R=18.9 \pm 0.3$ $\mu$m. The distance between pipettes, $\delta$, decreases over time as the aggregate is compressed. The crystal undergoes six well defined transitions $p \rightarrow (p-1)$ evidenced by six local maxima in the force curve. (b) Zoom of the $6 \rightarrow 5$ transition force peak, corresponding to the black square shown in panel (a). The right side of the peak, rising of the force, occurs during the compression of the crystal. The onset of the peak is shown by the black dot. The force reaches a maximum $F_{\textrm{c}}$, shown by the blue dot, as the crystal fractures.  (c-g) Sequence of microscopy images of the crystal being compressed (scale bar is 50~$\mu$m). During a fracture event, all bonds are broken in a single catastrophic and coordinated manner along fracture lines shown by the red dashed lines in (c). The fracture patterns for 2D crystals consist  of equilateral triangles  with $(p-1)$ droplets on the triangle's side [see (c), (f) and (g)].}
  \label{fig2}
\end{figure}

Figure~\ref{fig2}~(a) shows the measured force as a function of the distance between the pipettes, $\delta$, for a crystal with initial geometry ($p_\mathrm{ini}=7$; $q_\mathrm{ini}=7$). 
As the crystal is compressed, the distance $\delta$ decreases over time; thus, in the plots the experiment progresses from right to left as indicated by the time arrow in Fig.~\ref{fig2}~(a). The trace shows six distinct peaks corresponding to six fracture events (see movie M2 in Supplemental Material). To accommodate for the decreasing space between the pipettes, the number of rows of droplets, $p$, must be reduced. These breaking events are referred to as row-reduction transitions: from $p=7$ to $p=6$ rows of droplets [designated as $7 \rightarrow 6$ in Fig.~\ref{fig2}~(a)], followed by $6 \rightarrow 5$, $5 \rightarrow 4$, etc, until the last transition $2 \rightarrow 1$. A zoom of the second transition ($6 \rightarrow 5$) is shown in Fig.~\ref{fig2}~(b) along with an optical image sequence corresponding to this specific transition. To undergo a transition, adhesive bonds between droplets that arise from SDS-induced depletion forces must be broken. The droplet assembly maximizes the total number of bonds between droplets, creating a hexagonal structure. We find that bonds break in a catastrophic and coordinated manner. As the crystal transitions from $p$ to $(p-1)$ rows, bonds are broken along $60^{\circ}$ fracture lines with respect to the pipettes [red dashed lines on Fig.~\ref{fig2}~(c)], resulting in equilateral triangles with $(p-1)$ droplets on their side. This breaking pattern corresponds to the least number of broken bonds -- and thus the minimum energy cost -- for the aggregate to undergo the row reduction transition. A simple geometrical calculation, ignoring edge effects, shows that the number $b$ of broken bonds satisfies $b=2q$. 

The bond breaking events observed are T1 events during which four droplets exchange neighbours~\cite{Chen2015,Gai2016}. T1 events along a fracture line at $60^\circ$ have also been reported as monodisperse droplets flow through a tapered channel~\cite{Gai2016,Gai2019}. Similar observations for simulations of lightly repulsive colloids under compression have been made~\cite{McDermott2016}. After the fracture, the $(p-1)$-sided equilateral triangles slide past each other to rearrange into a new crystal with $(p-1)$ rows of droplets accommodated between the pushing and the force-sensing pipettes. During the sliding of the structure between two fracture events the force is nearly zero as can be seen in Fig.~\ref{fig2}~(b). This indicates that the compression is conducted at low enough speed (0.3~$\mu$m.s$^{-1}$) to ensure that the viscous drag is negligible. Nevertheless, the sharp decay of the force peak corresponds to the resolved relaxation of the force-sensing pipette as the entire system evolves through a viscous medium [Fig.~\ref{fig2}~(a) and (b)]. Separate experiments have shown that the relaxation of the force-sensing pipette in absence of droplets is almost instantaneous on the relevant timescale. Therefore, the decay of the force mentioned above is mainly due to the viscous damping experienced by the droplets, though it remains fast compared to the timescale of the experiment. For example, for the peak shown in Fig.~\ref{fig2}(b), the decay occurs over $~ 5$~s while the experiment takes~$10$~min (the frame rate is 1.8 Hz). The relaxation of the pipette takes longer for later transitions, for example the $2\rightarrow 1$ transition, as more droplets must flow over a larger distance. 

As a crystal aggregate is compressed, the force builds and the system is elastically deformed. The elastic energy stored eventually reaches the point of breaking the adhesive bonds. Another interesting feature of the force trace is the evolution of the peak magnitude with the transition index. From Fig.~\ref{fig1}~(a), it is clear that the force required to fracture the structure becomes larger from one transition to the next, as $p$ diminishes. This force increase can be explained by the number of bonds that need to be broken to enable the transition. In going from one transition to the next, the number of rows, $p$, decreases and thus $q$ increases ($N_\mathrm{tot}=pq$ is constant).  Since $q$ increases and the minimal number of broken bonds is given by $b=2q$, a larger force is required with each subsequent transition. 

\subsection{Equivalent spring model}
To quantitatively understand the evolution of the magnitude of the peaks for subsequent transitions, we developed a minimal analytical model.
Let us focus on the onset of fracture, for example the peak shown in Fig.~\ref{fig2}~(b). If droplets were strict hard spheres, one would expect the rise of the peak (\emph{i.e.} upon decreasing $\delta$) to have an infinite slope. The fact that the force increases with a finite slope indicates that droplets are slightly deformable and the aggregate can withstand a certain amount of elastic deformation before breaking. However, the deformation of individual droplets remains extremely small and cannot be seen on optical microscopy images (see movie M2 in Supplemental material). To account for this capillary deformation, we model individual droplets as identical springs with spring constant $k_1$. This is consistent with previous studies that have investigated the deformation of a droplet under external forces~\cite{pontani2012,Edwards2003}. At lowest order in the deformation, the restoring force can indeed be modelled as that of a capillary spring with a spring constant $k_{1}=\cal{G}\gamma$ (see discussion in the appendix) directly proportional to the interfacial tension $\gamma$ between oil and the SDS aqueous solution~\cite{Attard2001}. The proportionality constant $\cal{G}$ depends on the exact contact geometry between the droplets. The aggregates are thus made of $q$ springs in parallel (along the same row) and $p$ rows of springs in series. The resulting crystal can thus be represented by an equivalent spring with constant: 
\begin{equation}
k_\mathrm{eq}= k_1q/p\ . 
\label{spring}
\end{equation}
Under compression, the restoring force is linear with slope $k_\mathrm{eq}$. To validate this model, we measure the evolution of the equivalent spring constant of a crystal for the different transitions as the aspect ratio $q/p$ changes from one transition to another.

\begin{figure}[h]
\centering
  \includegraphics[width=.9\columnwidth]{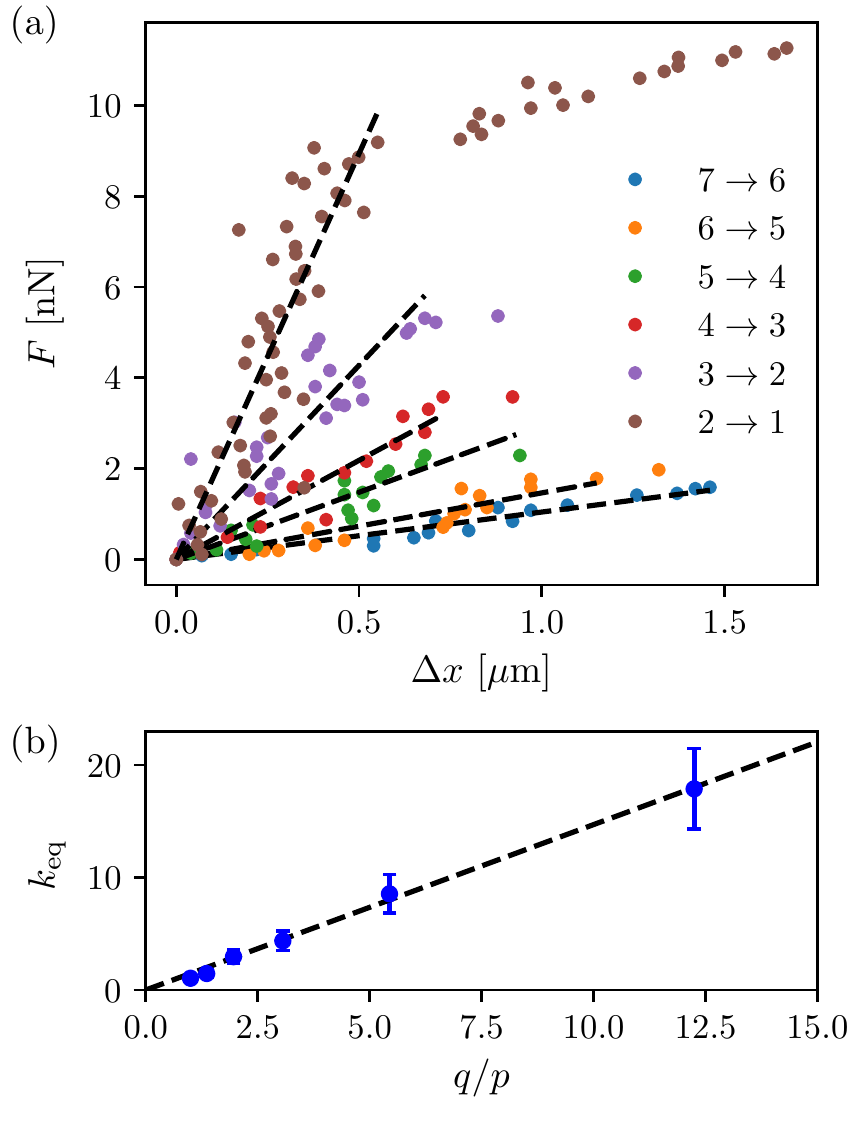}
    \caption{(a) Force as a function of the compression $\Delta x$ of the aggregate for the six different transitions. The part of the force shown is the rising one, that corresponds to the force between the black and blue dot in Fig.~\ref{fig2}~(b), from onset to peak. The compression is defined to be equal to zero at the onset. The black dashed lines are linear best fits to the data. Each dashed line is only fitted to the elastic part of the compression. The slopes of these lines are the equivalent spring constants of the cluster, $k_\mathrm{eq}$. (b) Evolution of $k_\mathrm{eq}$ with the aspect ratio $q/p$. The black dashed line is the best fit to Eq.~\ref{spring}. The error bars correspond to the uncertainty on the slopes of the linear fits in panel (a). Indeed, these linear fits are sensitive to the first and last data points included in the fit. }
  \label{fig3}
\end{figure}

    Figure~\ref{fig3}~(a) shows the rise of the force peaks as a function of the compression of the crystal for the six $p \rightarrow (p-1)$ transitions. For each peak, the compression is defined as $\Delta x= \delta_0^{\,p}-\delta$ with $ \delta_0^{p}$ the onset of compression for each transition, corresponding to the black dot in Fig.~\ref{fig2}~(b). Within the resolution of the experiment, the force is linear with the compression. The peaks flatten and deviate from the early linear behaviour for larger values of $\Delta x$. This is particularly noticeable on the curve corresponding to the $2 \rightarrow 1$ transition and can be explained by a slight misalignment of the pushing and force-sensing pipette. The misalignment results in parts of the crystal breaking earlier than the rest (movie M2 in Supplemental Material). The experiment is more sensitive to misalignment for the later transitions as the lateral extent of the crystal is larger (increasing number $q$ of droplets per rows). The part of the curve that deviates from the linear behaviour is excluded from the linear fit as the model is only valid in the limit where all droplets are in contact and prior to the onset of fracture events. The slope of each force curve $k_\mathrm{eq}$ is extracted and plotted against the ratio $q/p$ as suggested by Eq.~\ref{spring} and shown in Fig.~\ref{fig3}~(b). The equivalent spring constant $k_{\textrm{eq}}$ is found to scale linearly with $q/p$, as predicted by Eq.~\ref{spring}, with a slope corresponding to $k_1\approx1.46$~mN/m. The value of surface tension extracted for the relation $k_1=\cal{G} \gamma$ depends on the geometrical factor $\cal{G}$. Literature values for the interfacial tension are in the $\sim 5-10$ mN/m range~\cite{rehfeld1967,mason1995}, which is consistent with what we find for a geometrical pre-factor, $\cal{G}$, on the order of 1. The assembly of droplets thus behaves like a perfect 2D Hookean solid, where the 2D-equivalent applied stress $\sigma = F/(2qR)$ is equal to the strain $\epsilon\sim \Delta x/(2pR)$ times a 2D-equivalent Young's modulus $E\sim \cal{G}\gamma$. 
    
In the equivalent spring model, it is assumed that the droplets can store elastic energy under compression and the stretching of adhesive bonds is neglected. Indeed, the adhesion comes from the depletion forces induced by the SDS micelles and can be described by the Asakura-Oosawa potential~\cite{crocker1999}. This potential has a negative curvature which means that as soon as the adhesive force is overcome, the bond between droplets breaks completely. The range of the depletion forces is set by the nanometric size of the SDS micelles and is thus short-ranged compared to the elastic deformation of the droplets. 

Having validated the equivalent spring model, if the aggregate maintains a compressed hexagonal configuration the droplets deform, and the stored elastic energy increases with compression as $E_{\textrm{s}}= k_{\textrm{eq}}\Delta x^2/2$. In order for the cluster to fracture, bonds between droplets must be broken. As mentioned above, the minimum number of bonds that must break for a 2D hexagonal crystal to rearrange is $b=2q$.  As a result, crystals fracture when the stored elastic energy,  $k_\mathrm{eq} \Delta x_\mathrm{c}^2/2$, reaches the threshold $2qE_1$, where $E_1$ is the depletion-induced adhesive energy associated with breaking a single bond. This criterion corresponds to a critical yield force $F_{\textrm{c}}=k_\mathrm{eq}\Delta x_\mathrm{c}$, which is equal to:
\begin{equation}
    F_{\textrm{c}} = 2N_\mathrm{tot}\sqrt{\frac{k_1E_1}{p^3}}\ .
    \label{eq:maxforce}
\end{equation}

The model developed in this study assumes an ideal experiment where the pushing and the force-sensing pipettes are strictly parallel and the $b$ bonds are broken simultaneously for every transition. As such, Eq.~\ref{eq:maxforce} represents an upper bound for the measured real force. It can be tested experimentally by recording $F_{\textrm{c}}$, for each transition $p \rightarrow (p-1)$, and with aggregates of different initial geometries $(p_{\textrm{ini}}$, $q_{\textrm{ini}})$. For example, the compression experiment shown in Fig.~\ref{fig2}~(a) for droplets with radius $R=18.9 \pm 0.3 \ \mu$m, leads to six values of $F_{\textrm{c}}$ for $p=7$ to $p=2$. After a compression experiment is completed, the same droplets can be reassembled into a new crystal with a different initial geometry. By using the same droplets from one experiment to another, we ensure that the energy per bond $E_1$ and spring constant $k_1$ remain constant. 
\begin{figure}[h]
\centering
  \includegraphics[width=.9\columnwidth]{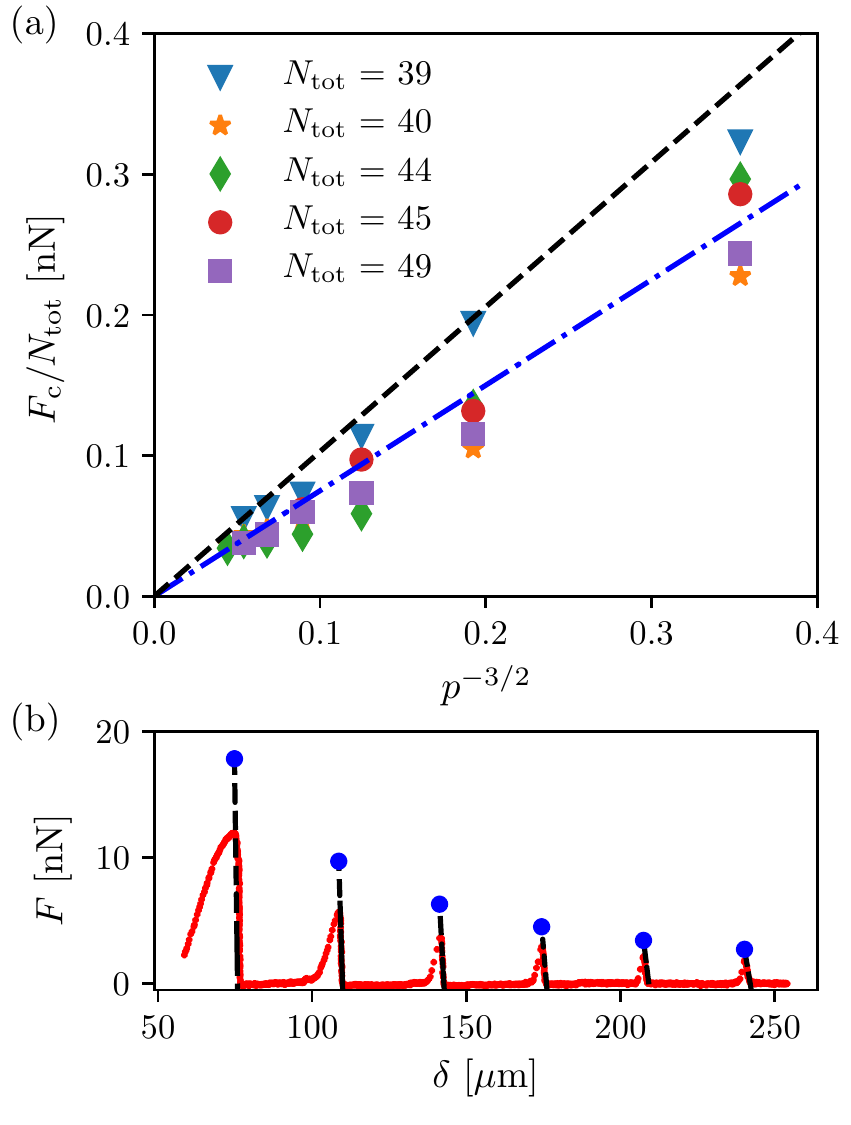}
  \caption{(a) Evolution of the peak force, $F_{\textrm{c}}$ normalized by $N_\mathrm{tot}$, as a function of $p^{-3/2}$ for five different compression experiments with different initial geometries. The size of the droplets is kept constant to $R=18.9 \pm 0.3$~$\mu$m. The initial geometries are:  (\protect\markertriangle) 7 alternating rows of 6 and 5 droplets; (\protect\markerstar) $p_\mathrm{ini}$=5; $q_\mathrm{ini}$=8;  (\protect\markerdiamond) 8 alternating rows of 6 and 5 droplets;  (\protect\markercirc) $p_\mathrm{ini}$=5; $q_\mathrm{ini}$=9; (\protect\markersquare) $p_\mathrm{ini}$=7; $q_\mathrm{ini}$=7. The dash-dot line corresponds to the best fit of Eq.~\ref{eq:maxforce} to all the data with the energy per bond $E_1= 0.096$~fJ, while the dashed line ensures all measured maximal forces fall below the line with $E_1\approx  0.2$~fJ. (b) Measured force as a function of the distance between pipettes for a ($p_\mathrm{ini}$=7; $q_\mathrm{ini}$=7) crystal. The blue data points correspond to the maximal forces, predicted from Eq.~\ref{eq:maxforce} with $E_1\approx  0.2$~fJ, which would be expected for an ideal experiment. The black dashed lines correspond to the theoretical predictions from Eq.~\ref{spring} with $k_1\approx1.46$~mN/m.}
  \label{fig4}
\end{figure}

In Fig.~\ref{fig4}~(a), we plot the experimentally measured force at failure normalized by $N_{\textrm{tot}}$ as a function of $p^{-3/2}$ for five different experiments with different initial geometries, as suggested by Eq.~\ref{eq:maxforce}. The best fit of  Eq.~\ref{eq:maxforce} to all the data (dot-dash line) corresponds to a bond energy of $E_1= 0.096$~fJ. 
The depletion energy per unit area that must be overcome to break an adhesive bond can be expressed as $W=\rho kTa$~\cite{israelachvili2011}, with $\rho\sim5.10^{23}$~m$^{-3}$ the number concentration of SDS micelles, $a\sim 5$~nm the radius of a micelle and $k$ the Boltzmann constant. With an estimate of the contact patch to be $R_\mathrm{p}\sim 0.1 R$, one can obtain the order of the energy per unit bond to be $E_1= W \pi R_\mathrm{p}^2\sim \rho kT a \pi R_\mathrm{p}^2 \sim0.13$~fJ, an approximate value that is  consistent with the best fit value. However, any imperfections in the experiment or thermal fluctuations, ensure that one can never measure a force greater than that corresponding to the ideal crystal. For comparison we also show Eq.~\ref{eq:maxforce} with $E_1= 0.2$~fJ (dashed line) corresponding to a bond energy adjusted such that all the data lies below the upper bound given by the theory. 
Using the experimental estimated value of $E_1\approx 0.2$~fJ, and the spring constant $k_1\approx1.46$~nN/m of individual droplets, one can construct the theoretical force curve for an ideal system through the various transitions. Figure~\ref{fig4}(b) shows the force trace for a ($p_\mathrm{ini}$=7; $q_\mathrm{ini}$=7) crystal (red curve). Equation~\ref{eq:maxforce} predicts the upper-bound values of the force peaks which are shown with the blue dots. Additionally, from Eq.~\ref{spring} we have that upon compression the force rises linearly with slope $k_\mathrm{eq}\approx (1.46\, q/p)$~mN/m, which is shown with the black dashed lines for the various transitions. We find good agreement between the simple analytical model proposed in this study and the experimental results. The discrepancy between the measured and the predicted maximum forces is larger for the later transitions (for example $2 \rightarrow 1$). This is expected since the experiment is more sensitive to imperfections as the lateral extent of the crystal becomes larger (increasing $q$).

\section{Conclusions}

In summary, we prepared 2D defect-free colloidal mono-crystals by assembling highly monodisperse droplets into small size aggregates. The force trace measured during compression shows a well defined number of peaks corresponding to row-reduction transitions. Using our experimental apparatus, we are able to measure macroscopic mechanical properties while monitoring individual droplets. Under small applied forces, crystals respond elastically. A simple assembly of capillary springs, representing individual droplets, captures the elastic properties of the crystal. As the aggregate is further compressed and a critical yield force is reached, the crystal fails catastrophically, but in a coordinated manner. Plastic T1 events occur simultaneously along $60^\circ$ fracture lines, resulting in equilateral triangles of droplets which slide past each other. The droplets eventually reassemble into a new crystal with one less row. An analytical model balancing the stored elastic energy upon compression with the released depletion-induced adhesion energy during bond breaking, allows us to predict the yield point. While here the attraction between droplets is caused by depletion forces, the model is expected  to remain valid for other sources of adhesion. The low-polydispersity droplet system with controllable adhesion strength provides an ideal platform for investigating material properties while individual constituents can be directly imaged. 

\section*{Conflicts of interest}
 There are no conflicts to declare.

\section*{Acknowledgements}
We gratefully acknowledge financial support by the Natural Science and Engineering Research Council of Canada. The work of E.R.W. was supported by the National Science Foundation (CBET-1804186). We also thank Maxence Arutkin for preliminary discussions.

\section*{Appendix}

\subsection*{Spring constant of a droplet}

\begin{figure}[h]
\centering
  \includegraphics[width=.9\columnwidth]{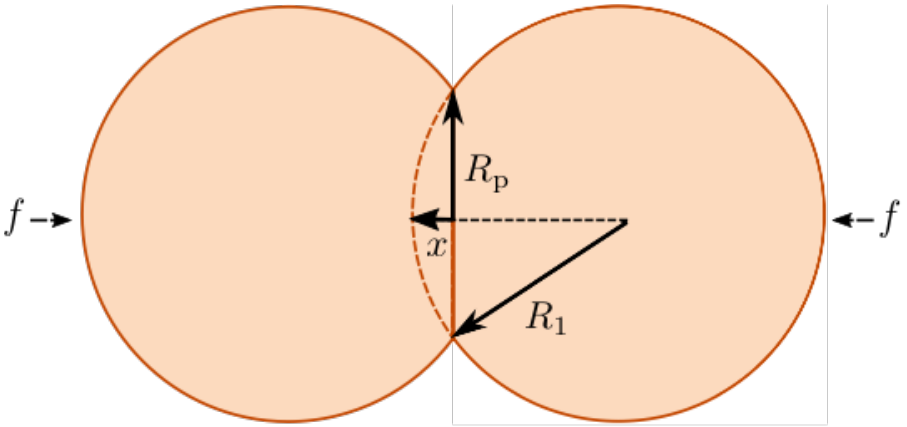}
  \caption{Schematic of two oil droplets being pushed against one another with force $f$. The droplets deform by an amount $x$ during compression and a patch with radius $R_{\textrm{p}}$ is formed between the two droplets. The volume of the droplet is conserved during compression and thus the initial radius, $R$, increases to a different value, $R_1$.}
  \label{fig5}
\end{figure}

Individual droplets are modeled as capillary springs of constant $k_1$, directly proportional to the surface tension $\gamma$ between the oil and the aqueous solution. In this section, we provide details to justify this model analytically. Consider two oil droplets, with initial radius $R$, compressed against one another as shown in Fig.~\ref{fig5}. 
During compression, the volume of the droplet is conserved but the radius, $R_1$, and thus the surface area of the droplet is modified. The change in surface area results in a capillary energetic cost. Following the work by Pontani {\it et al.}~\cite{pontani2012}, one can show that the change in energy is $\Delta E =  \gamma \pi R^2\theta^4/2$. For the spring model developed in this study, the change in energy must be expressed as a function of the compression of an individual droplet, $x$. Using simple geometrical arguments, the compression can be written as~\cite{israelachvili2011} $x\simeq R_{\textrm{p}}^2/R$, with $R_{\textrm{p}}$ the radius of the patch between droplets (see Fig.~\ref{fig5}). In the limit of small deformations: $\theta\simeq R_{\textrm{p}}/R$. Using these two expressions, the change in surface energy can be written as:
\begin{equation}
\Delta E\simeq {1\over 2} \gamma \pi x^2\ .
\label{eq:springmodel}
\end{equation}
Comparing Eq.~\ref{eq:springmodel} to the energy $E_{\textrm{s}}=k_1x^2/2$ stored in a spring naturally leads to $k_1=\pi \gamma$. Taking into account the hexagonal geometry of our droplet assemblies would lead to a different pre-factor for the effective individual spring constant $k_1=\cal{G}\gamma$. Nevertheless, this simple analytical calculation justifies why droplets can be modeled as springs, and that the associated spring constant is directly proportional to the surface tension $\gamma$. 





\bibliography{Onoditbiot2020} 
\bibliographystyle{rsc} 

\end{document}